\documentclass[aps,pra,showpacs,groupedaddress]{revtex4}
\usepackage{graphicx}
\begin{document}
\title{Optical squeezing of a mechanical oscillator by dispersive interaction}
\author{M Bhattacharya, P -L  Giscard and P Meystre}
\address{B2 Institute, Department of Physics and College
of Optical Sciences, The University of Arizona, Tucson,
AZ 85721, USA}
\begin{abstract}
We consider a small partially reflecting vibrating mirror
coupled dispersively to a single optical mode of a high
finesse cavity. We show this arrangement can be used to
implement quantum squeezing of the mechanically
oscillating mirror.
\end{abstract}
\pacs{07.10.Cm, 42.50.Pq, 06.30.Bp, 04.80.Nn}
\maketitle
\section{Introduction}
Experimental advances in nanofabrication and in laser cooling and
trapping have turned optomechanical systems into viable
laboratories for the observation of quantum mechanics at
macroscopic scales. Non-equilibrium cooling of small movable
mirrors using laser-driven cavities has been demonstrated
experimentally by a number of groups
\cite{pinard1999,karrai2004,gigan2006,kleckner2006,arcizet2006,
schliesser2006,corbitt2007,rugar2007}. Likewise, theory has shown
that in principle these methods should be able to lower the mirror
to its quantum mechanical ground state
\cite{tombesi2003,mishpm1,mishpm2,kippenberg2007,marquardt2007}.
The preparation of that state of the mirror is an important first
step in exploring characteristic features of quantum mechanics
such as superposition \cite{leggett2002} and entanglement in
macroscopic systems \cite{mancini2002}.

Squeezed states have also attracted much attention, due to their
favorable quantum noise properties \cite{mankobook}. Squeezed
states of light are expected to find applications in precision
measurements \cite{caves1981} and optical communications
\cite{takahashi1965,yuen1976}. In a parallel development, the
squeezing of classical noise in mechanical oscillators has been
demonstrated in optomechanical cavities \cite{heidmann2003}, ion
traps \cite{pritchard1992,pritchard1995} optical lattices
\cite{rolston1997} and other systems \cite{rugar1991,rugar1993}.
Quantum squeezing of phonons has been achieved in ion traps
\cite{wineland1996} and in crystals \cite{nori1997,merlin1997}.
Proposals to realize squeezed states of nanomechanical oscillators
in the quantum regime have been made involving two-mirror cavities
\cite{tombesi2002}, parametric mixing in solid state circuits
\cite{long2007,nori2007,simmonds2006}, microwave coupling to a
charge qubit \cite{zoller2004}, quantum non-demolition
measurements \cite{korotkov2005} and the parametric modulation of
a mechanical spring \cite{wybourne2000}. Their application to
gravitational interferometry has also been discussed
\cite{hollenhorst1979}. Other nonclassical states such as
Schrodinger `cats' have been proposed using movable cavity mirrors
\cite{knight1997}.

This article shows how to realize a squeezed state of a
mechanically moving mirror in a high finesse optical cavity.
Previous proposals to achieve this goal have relied on the
mathematical analogy between an optical resonator with a moving
mirror and a Kerr medium, and the mechanism of squeezing has been
\textit{parametric driving}. Here we invoke \textit{compression}
as an alternative route to squeezing \cite{note1}. In that scheme
squeezing of the mirror motion relies on coupling it dispersively
with the cavity, a possibility that has recently been pointed out
\cite{thompson2007} and analyzed in detail \cite{bum2007}. To
provide a complete discussion we consider not one but two modes of
the cavity, the moving mirror being coupled dispersively to one of
the modes and dissipatively to the other. This configuration was
recently proposed as an efficient cooling and trapping
configuration for semi-transparent mirrors \cite{bum2007}; here we
show that this configuration also allows for displacing and
squeezing the error ellipse of the oscillating mirror in phase
space.

The remainder of the paper is organized as follows.
Section~\ref{sec:Ham} introduces the physical system and its model
Hamiltonian, section~\ref{sec:Ev} discusses the corresponding
evolution operator and the resulting displacement and squeezing
assuming that the moving mirror starts from its quantum mechanical
ground state. Section~\ref{sec:therm} discusses the effects of
squeezing in the presence of noise and damping. 
Section~\ref{sec:Con} supplies a conclusion and an outlook.

\section{The Hamiltonian}
\label{sec:Ham}

We consider a high finesse cavity with two perfectly reflecting
fixed end mirrors, and a partially reflective movable middle
mirror as shown in Fig.1. 
\begin{figure}
\includegraphics[width=0.45 \textwidth]{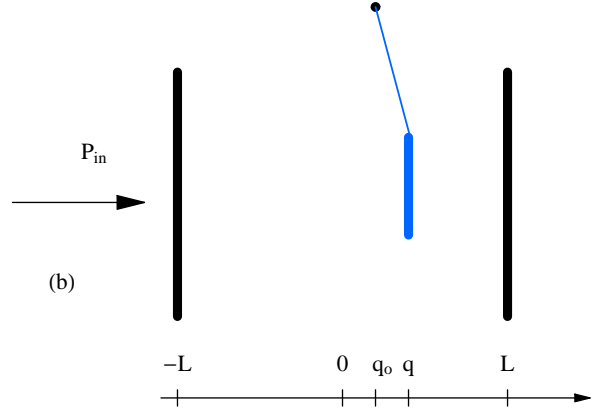}
\caption{\label{fig:tmc} The three-mirror cavity arrangement. 
The two outer mirrors are fixed and the middle mirror vibrates 
harmonically about its equilibrium position $q_{o}$. 
$P_{in}$ is the laser power coupling into the cavity.}
\end{figure}
The middle mirror is assumed to execute
small harmonic oscillations $q$ about its equilibrium position.
It couples dissipatively (linearly in $q$) to a
cavity mode $a$ of frequency $\omega_D$ and dispersively
(quadratically in $q$) to a second mode $b$ of frequency
$\omega_S$. The Hamiltonian $H'$ modelling the system is derived
in Ref.~\cite{bum2007}, and is given explicitly by
\begin{eqnarray}
\label{eq:Hq}
H' =\hbar \omega_D (a^{\dagger} a+\frac{1}{2}) +
\hbar \omega_S (b^{\dagger} b +\frac{1}{2})+
\frac{p^2}{2m}+\frac{1}{2}m \omega_{m}^{2}q^2+\hbar
\xi_Da^{\dagger}a q +\hbar \xi_Sb^{\dagger}b q^2,
\end{eqnarray}
where $\omega_m$ is the oscillation frequency of the middle
mirror,
\begin{equation}
 |\xi_D|=\frac{\sin2k_{n}q_{0}}{\sqrt{(1-T)^{-1}-\cos^{2}2k_{n}q_{0}}}\xi,
\end{equation}
with $q_0$ the equilibrium position of the moving mirror of
transmissivity $T$, $\omega_n=n\pi c/L$, $\xi=\omega_n /L$ and
$k_n=\omega_n /c$, and
\begin{equation}
|\xi_S|=\frac{\tau\xi^{2}}{2}\left(\frac{1-T}{T}\right)^{1/2},
\end{equation}
where $\tau = 2L/c$.
The frequencies of the modes $a$ and $b$ can be chosen such that
$\xi_{D,S}$ are either positive or negative. In the case of
$\xi_{S}$ this corresponds to the use of trapping and
anti-trapping modes, respectively \cite{bum2007}.

For $\xi_D <0$ we have
\begin{equation}
 \omega_D = \omega_n-\frac{1}{\tau}\left[\sin^{-1}\left(\sqrt{1-T}
\right)-\sin^{-1}\left(\sqrt{1-T}\cos 2k_{n}q_{0}\right)\right],
\end{equation}
and for $\xi_D >0$
\begin{equation}
 \omega_D = \omega_n+\frac{\pi}{\tau}-\frac{1}{\tau}\left[\sin^{-1}
\left(\sqrt{1-T}\right)+\sin^{-1}\left(\sqrt{1-T}\cos 2k_{n}q_{0}
\right)\right].
\end{equation}
Similarly,
\begin{equation}
\omega_S=\omega_n
\end{equation}
for $\xi_S <0$ and
\begin{equation}
\omega_S=\omega_n+\frac{2}{\tau}\cos^{-1}(1-T)^{1/2}.
\end{equation}
for $\xi_S >0$.

The first two terms in the Hamiltonian $H'$ describe the energies
of the optical modes, the next two the energy of the oscillating
mirror, and the last two the dissipative and dispersive coupling
energies. The bosonic modes obey the commutation relations
$[a,a^{\dagger}]=1$ and $[b,b^{\dagger}]=1$, and the dynamical
variables of the oscillating mirror follow the canonical
commutation relation $[q,p]=i\hbar$.

The Hamiltonian~(\ref{eq:Hq}) indicates that for low values of
$\xi_S$ the spring potential energy dominates the anti-trapping
due to radiation pressure, hence the middle mirror still behaves
as a harmonic oscillator, but of lower frequency. However, for
$\xi_s <0$ increasing $|\xi_S|$ leads to a point 
\begin{equation}
\label{eq:crit}
C_S=-\omega_m/2,
\end{equation}
where the mirror behaves as a free particle. For even higher 
values of $|\xi_S|$ radiation pressure-induced anti-trapping 
dominates and the mirror behaves like an inverted harmonic 
oscillator \cite{lo1991}. We do not consider that regime in this 
paper. This is consistent with the assumption of small mirror 
displacements $q$ used to derive the Hamiltonian~(\ref{eq:Hsc}), 
as well as with requirements of stability.

In the following we consider a semiclassical version of the
Hamiltonian $H'$ valid for situations where the optical modes can
be treated classically. In that case
\begin{equation}
\label{eq:semic}
\label{eq:cnos} a \rightarrow \alpha, \hspace{0.2in} b \rightarrow
\beta,
\end{equation}
and expressing the mirror displacement in terms of raising and
lowering operators
\begin{equation}
q=\sqrt{\frac{\hbar}{2m\omega_{m}}}(c^{\dagger}+c),
\end{equation}
we have
\begin{equation}
\label{eq:Hsc} H' \rightarrow H=\hbar C_D (c + c^\dagger)+2\hbar
C_R K_{0}+\hbar C_S(K_{-}+K_{+}).
\end{equation}
where we have removed a constant energy $E_0=\omega_D
(|\alpha|^2+\frac{1}{2})+\omega_S(|\beta|^2+\frac{1}{2}),$ and
\begin{eqnarray}
\label{eq:Hconstants}
C_D&=&\frac{\xi_D|\alpha|^2}{\sqrt{2m\omega_m/\hbar}}, \nonumber \\
C_S&=&\frac{\hbar \xi_S |\beta|^2}{m \omega_m}, \nonumber \\
C_R&=& C_S+\omega_m.
\end{eqnarray}
In the semiclassical Hamiltonian $H$ we have also introduced the
operators \begin{equation}
    K_{0}=(c^\dagger c+c c^\dagger)/4,\,\,\,\,\,K_-=c^2/2\,\,\,\,,\,K_+=c^{\dagger2}/2,
\end{equation}
which together with $c$ and $c^\dagger$ form the basis of the
so-called two-photon Lie algebra \cite{wunsche2002}, with
\begin{eqnarray}
\label{eq:commrules}
    \left [K_0,K_\pm \right ]&=& \pm K_\pm , \,\,\,\,
    \left [K_-, K_+\right ]= 2 K_0,  \nonumber \\
    \left [K_-,c \right ]&=& \left [K_+,c^\dagger\right ]=0, \nonumber \\
    \left [K_-,c^\dagger \right ]&=& c, \,\,\,\,
    \left [K_0,c^\dagger \right ]=c^\dagger/2.
\end{eqnarray}
As is well known, the operators $\{c, c^\dagger\}$ and $\{K_0,
K_\pm \}$ form two sub-algebras, the associated operators forming
the generators of coherent states and of squeezed states,
respectively. The Hamiltonian~(\ref{eq:Hsc}) has previously been
studied in some detail in the context of molecular
translational-vibrational interactions \cite{yuan1987} and
laser-plasma scattering \cite{mann1985} and very recently in the 
context of atomic vapors inside resonators \cite{norip2007}. See also
\cite{mandelbook,fernandez1989} for additional discussions of this
model.

\section{Time evolution}
\label{sec:Ev}

\begin{figure}
\includegraphics[width=0.90 \textwidth]{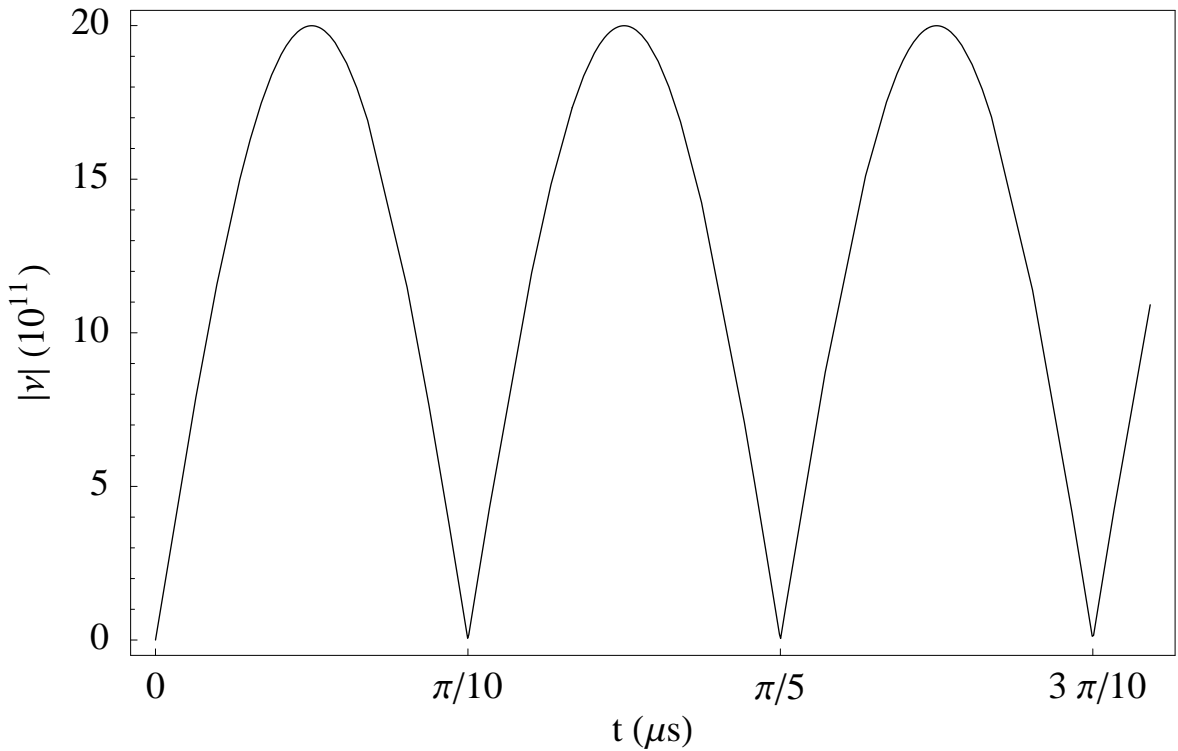}
\caption{\label{fig:displace}The modulus of the dimensionless
displacement amplitude $|\nu|$ defined using
Eq.~(\ref{eq:complexnu}) as a function of time.
The parameters used are cavity length $L=5$mm, laser wavelength
$\lambda =514$nm, and a middle mirror of mass $m=1\mu$g, 
vibration frequency $\omega_m=2\pi2.5$kHz, damping constant 
$D_m=0.02\mu$gHz, transmissivity $T=10^{-4}$, equilibrium position
$\lambda/10$ and base temperature $T_e=100$mK. The end mirror 
transmissivity has been taken to be $10^{-5}$ and the power
coupling into the mode 1mW.}
\end{figure}

Using the Lie-algebraic symmetries of $H$, the associated
evolution operator can be disentangled as \cite{wunsche2002}
\begin{equation}
\label{eq:U}
 U=\exp[-iHt/\hbar]=e^{i\delta} D(\nu)R(\phi)S(\kappa),
\end{equation}
where $\delta$ is an unimportant overall phase, and
\begin{equation}
 D(\nu)=e^{\nu c^\dagger-\nu^* c},
\end{equation}
is a displacement operator, with \cite{wunsche2002}
\begin{equation}
    \label{eq:complexnu} \nu =
    \frac{C_D}{\chi}\left[\frac{\omega_m}{\chi}\left (\cos\chi
    t-1\right)-i\sin \chi t \right]
\end{equation}
and
\begin{equation}
    \label{eq:chi}
    \chi = \sqrt{C_R^2-C_s^2}=\left[\omega_m(\omega_m + 2 C_S)\right]^{1/2}.
\end{equation}
In the bound oscillator regime, i.e. for $C_S > - \omega_m/2$,
we have $\chi^2>0$, and we can choose $\chi>0$ without 
loss of generality. That parameter largely determines 
the time scale of the mirror dynamics; in the absence 
of squeezing $(C_S=0)$ it is just the harmonic oscillator 
period. The factor in parentheses in Eq.~(\ref{eq:chi}) 
quantifies the mismatch from the condition 
$C_S = - \omega_m/2$ [Eq.~\ref{eq:crit}] which demarcates the 
regimes of qualitatively different physical behaviors in the 
system. From Eqs.~(\ref{eq:complexnu}) and ~(\ref{eq:chi}) 
$C_S$ can both increase or decrease the characteristic time 
scale of the displacement as well as its magnitude. The 
displacement in phase space is given by the absolute value of 
$\nu$. A plot of $|\nu|$ versus time for typical experimental 
parameters is shown in Fig.\ref{fig:displace}.
Near the first minimum i.e. for
\begin{equation}
\label{eq:timescale} t \ll 1/\chi,
\end{equation}
the displacement is linear in time to lowest order, i.e.
\begin{equation}
 |\nu| \simeq \left|C_D \left[ t-\frac{\omega_m}{3}\left(\frac{\omega_m}
{8}+C_S \right)t^3 \right] +\mathcal{O}\left[t^5\right] \right|,
\end{equation}
and the effects of squeezing come in at third order.
Interestingly, by adjusting the squeezing such that
$C_S=-\omega_m/8$, which is still in the $\chi^2 >0$ regime, 
the third-order time dependence of the displacement can be
removed. Qualitatively similar behavior can be seen near 
every minimum in Fig.~\ref{fig:displace}. We note that in 
the absence of squeezing $(C_S=0)$,
\begin{equation}
\label{eq:ssq}
 |\nu| \simeq \left|\frac{2 C_D}{\omega_m}\sin\frac{\omega_m t}{2} \right|,
\end{equation}
while for large squeezing $(\omega_m/\chi \ll 1)$,
\begin{equation}
\label{eq:ss} |\nu| \simeq \left|\frac{2 C_D}{\chi}\sin \chi t \right|
\end{equation}
For typical parameters we have $|\frac{2 C_D}{\omega_m}| \sim
10^{9},|\frac{2 C_D}{\chi}| \sim 10^{11}$. Therefore for 
both small and large squeezing a coherent mechanical state 
of the middle mirror of relatively large amplitude can be 
produced starting from the ground state.

Returning to the various components of the evolution operator
$U(t)$ we observe that
\begin{equation}
\label{eq:rot}
    R(\phi)=e^{i\phi K_0}
\end{equation}
is a rotation operator, and
\begin{equation}
\label{eq:skappa} S(\kappa)= e^{\kappa^{*} K_--\kappa K_+},
\end{equation}
is a squeezing operator, with
\begin{equation}
    \label{eq:modkappa}
    |\kappa|=\left|\sinh^{-1}
    \left(\frac{C_S}{\chi} \sin \chi t \right)\right|.
\end{equation}
As expected that operator does not depend on $C_D$, i.e. 
displacement does not affect squeezing.

It turns out that the rotation angle $\phi$ in Eq.~(\ref{eq:rot}) 
is exactly opposite the angle at which the squeeze operator 
tilts the error ellipse of the moving mirror in phase space 
\cite{mandelbook}, i.e.
\begin{equation}
\phi = - \left[\frac{\mathrm{phase}(\kappa)+\pi}{2} \right].
\end{equation}
The two rotations therefore cancel each other out and $\phi$ 
effectively plays no role in the dynamics. It is in fact 
intuitively clear that the effects of rotation should cancel 
out, i.e. the axes of the final error ellipse should be 
aligned along $p$ and $q$ in phase space. This is because 
Eq.~(\ref{eq:Hq}) stipulates that position is the only 
quadrature of the oscillating middle mirror that can be 
squeezed or anti-squeezed, the latter situation corresponding 
to momentum squeezing.

\begin{figure}
\includegraphics[width=0.85 \textwidth]{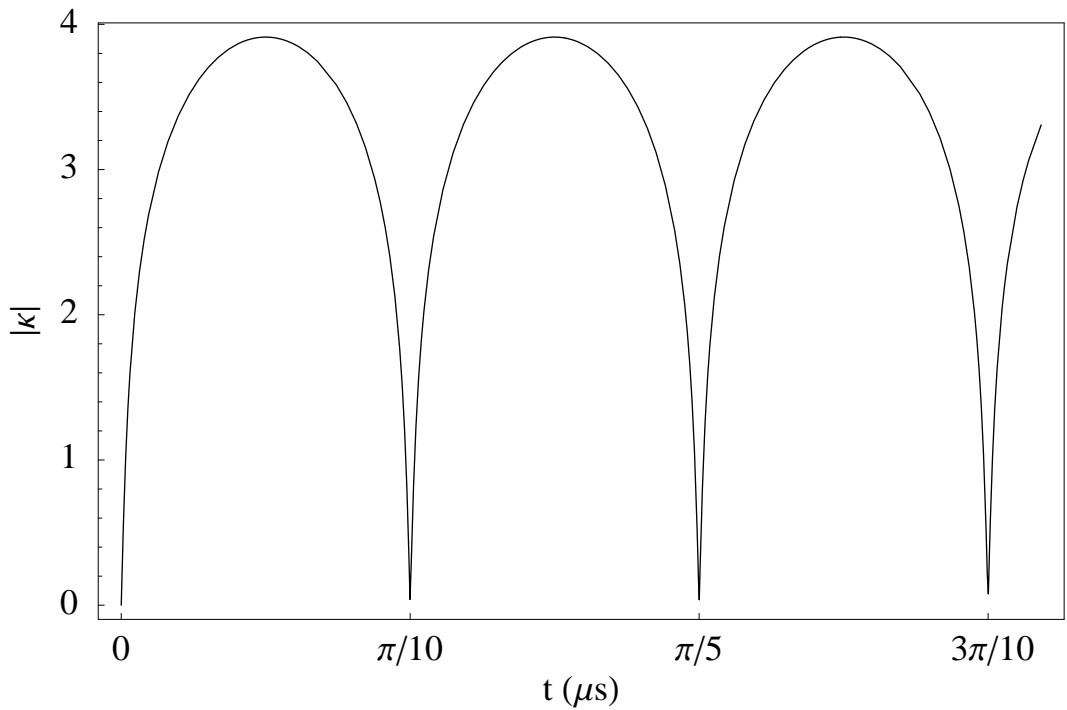}
\caption{\label{fig:squeeze}The modulus of $|\kappa|$ from
Eq.~(\ref{eq:modkappa}) as a function of time. The parameters 
are the same as in the caption of Fig.~\ref{fig:displace}.}
\end{figure}

Figure~\ref{fig:squeeze} shows $|\kappa|$ versus time 
for typical experimental parameters .
As can be seen from that plot the squeezing first grows 
linearly in time. This can be confirmed by analytically 
expanding Eq.~\ref{eq:modkappa} for the case of $t \ll 1/\chi$
\begin{equation}
    |\kappa| \simeq \left|C_S \left[t-\frac{(\omega_m+C_S)^{2}}{6}t^3
    \right]+\mathcal{O}\left[t^5\right]\right|.
\end{equation}
We note that the third-order time dependence can be removed for
$C_S=-\omega_m$. Actually under this condition it can readily be 
seen from Eq.~(\ref{eq:modkappa}) that all higher orders 
vanish and squeezing is purely linear in time : 
$|\kappa|=\omega_m t$. However that case corresponds 
to $\chi^2 <0$, a situation where the mirror does not behave
as a bound harmonic oscillator.

\section{Squeezing of thermal states}
\label{sec:therm}
If the middle mirror is prepared in its quantum mechanical ground
state, the squeezing operator~(\ref{eq:skappa}) produces a
squeezed vacuum \cite{scullybook}. From Fig.~\ref{fig:squeeze},
the maximum value of $\kappa$ is approximately 4, which implies a
maximum squeezing of $R=e^{-4} \sim 0.018$, or Log$_{10}(0.018)
\sim 18$dB of squeezing.

However the placement of a macroscopic nano-oscillator in its
ground state has not yet been achieved experimentally, so we also
consider thermal states of the middle mirror. They are characterized
by a thermal phonon number given by the Bose distribution
\begin{equation}
 n_T=\left[\mathrm{Exp}\left({\frac{\hbar \omega_m}{k_B T_e}}\right)-1\right]^{-1},
\end{equation}
where $T_e$ is the mirror equilibrium temperature and $k_B$ is 
Boltzmann's constant. Any realistic model should also include 
the damping of the mirror. We estimate these effects by including 
noise and damping in the Heisenberg equations of the mirror in a 
manner consistent with the fluctuation-dissipation theorem. This 
produces the corresponding quantum Langevin equations from 
Eq.~(\ref{eq:Hq}) in a standard way. Setting $\alpha=0$ for 
simplicity and concentrating therefore solely on the squeezing 
part of the Hamiltonian $H$ the quantum Langevin equations turn 
out to be 
\begin{eqnarray}
\label{eq:QLE}
\dot{q}&=&p/m, \nonumber \\
\dot{p}&=&-m\omega_m \chi^2 q-\frac{D_{m}}{m} p+\epsilon (t),
\end{eqnarray}
where $D_m$ is the damping constant of the mirror and
$\epsilon (t)$ represents Brownian noise with average zero
and fluctuations correlated as
\begin{eqnarray}
\label{eq:Brownian}
\langle \delta \epsilon (t) \delta \epsilon(t') \rangle=
D_{m}\int_{-\infty}^{\infty} \frac{d\omega}{2\pi}e^{-i\omega(t-t')}\hbar \omega
\left[1+ \coth \left(\frac{\hbar \omega}{2k_{\rm B}T_e}\right)\right]. \nonumber \\
\end{eqnarray}
For a high mechanical quality factor, the Brownian force becomes
delta-correlated in the time domain \cite{vitali2007}. In Fourier 
space the correlation can then be written as
\begin{equation}
\label{eq:fcorr}
\langle \delta \epsilon (\omega) \delta \epsilon (\omega') \rangle=
2 D_m \hbar \omega_m (2n_T +1)\delta (\omega+\omega').
\end{equation}
By setting the time derivatives equal to zero the steady-state
solutions to Eq.~(\ref{eq:QLE}) can easily be found to be
\begin{equation}
 q_s=p_s=0.
\end{equation}
Linearizing all operators in Eq.~(\ref{eq:QLE}) as sums of a
semiclassical steady-state value and a small quantum fluctuation
(i.e. $q=q_s+\delta q$) we obtain linear dynamical equations for
the fluctuations. Using Fourier transforms and
Eq.~(\ref{eq:fcorr}) we can solve the fluctuation equations to obtain
$\delta q (\omega)$, etc. We can therefore also find the (equal-time) 
correlation function for the position 
\begin{equation}
\label{fluctuations}
    \langle \delta q^2 \rangle = (2n_T+1)\frac{\hbar
    \omega_m}{2 m \chi^2},
\end{equation}
which is independent of time since the noise process we have 
considered is stationary [Eq.~(\ref{eq:Brownian})]. This result 
for the position uncertainty has followed from a linear response 
analysis, however it agrees to first order with results from more 
sophisticated computations \cite{weiss1984}. For example in the 
absence of squeezing $(C_S=0)$, and at high temperatures, 
$(n_T \sim k_{B} T_{e}/\hbar \omega_m \gg1 )$, 
\begin{equation}
\label{eq:hot}
\langle \delta q^2 \rangle =k_{B}T_{e}/m \omega_{m}^{2}.
\end{equation}
On the other hand for $C_S=0$ and low temperatures $(n_T \ll 1)$,
\begin{equation}
\label{eq:cold}
\langle \delta q^2 \rangle =\hbar/2m\omega_m,
\end{equation}
which is just the square of the oscillator length of the ground 
state of the moving mirror. The results in Eqs.~(\ref{eq:hot}) 
and ~(\ref{eq:cold}) agree with an earlier and more rigorous
derivation \cite{weiss1984}. Using Eq.~(\ref{fluctuations}) 
in the presence of squeezing $(C_S \neq 0)$ and defining a 
position uncertainty $R$ in terms of the ground state 
oscillator length we find
\begin{equation}
 R =\frac{\langle \delta q^2 \rangle^{1/2}}{\sqrt{\hbar/2m\omega_m}}
=\left[(2n_T +1)\frac{\omega_m}{\omega_m+2C_S}\right]^{1/2}
\sim \left(\frac{k_B T_{e}}{\hbar C_S} \right)^{1/2},
\end{equation}
where the last expression has been written in the limit of
high temperature and high squeezing. $R$ needs to be lower 
than 1 for squeezing to be present, i.e. the fluctuations
in the position need to be smaller than the ground state
uncertainty in position. For milliKelvin temperatures and 
hundreds of milliwatts of laser power, $R=0.15$ and about 
8dB of mechanical squeezing can be obtained, which is still 
considerable.

\section{Conclusion}
\label{sec:Con}
In conclusion we have considered a partially reflective
vibrating mirror coupled dispersively to an optical
mode of a high finesse cavity. We have shown that 
quantum squeezing of the mechanical motion of the mirror 
can be achieved in this way. We have described the unitary 
dynamics of the oscillator in some detail and shown that 
the squeezing remains non-negligible in the presence of 
noise and damping. Clearly the squeezing field itself can 
be employed in a time-dependent fashion although we have 
not investigated such a scenario.

It was also shown that the oscillator can be displaced 
by a second field to which it is coupled dissipatively. 
The dynamics of the displacement can be influenced by 
the squeezing field, although the converse is not true. 

It would be interesting to explore the effects of the fully
quantum mechanical Hamiltonian [Eq.~(\ref{eq:Hq})] without 
making the semiclassical approximation of Eq.~(\ref{eq:semic}). 
This may lead to highly non-classical states of the 
mirror-field system as found in the case of purely dissipative
coupling \cite{knight1997}. We are currently also working on 
generalizing the present proposal to the case of multiple 
mirrors in the same cavity.

\section{Acknowledgements}
This work is supported in part by the US Office of Naval
Research, by the National Science Foundation and by the 
US Army Research Office. We thank H. Uys and O. Dutta 
for useful discussions.

\end{document}